\documentclass[
 aip,amssymb,jcp,
reprint,floatfix,
]{revtex4-2}
\usepackage{graphicx}
\usepackage{color}
\usepackage{hyperref, url}
\usepackage{float}
\usepackage{array, multirow}
\usepackage{amsmath, amssymb, gensymb}
\usepackage{physics, dsfont}
\usepackage{newtxtext}
\usepackage{booktabs}
\usepackage{xcolor, soul}
\usepackage[normalem]{ulem}
\usepackage{leftindex}
\usepackage[T1]{fontenc}
\usepackage[caption=false]{subfig}
\usepackage{mathrsfs} 
\usepackage{array}
\hypersetup{colorlinks,
    linkcolor={blue!75!black!80!yellow},
    citecolor={blue!75!black!80!yellow},
    urlcolor={blue!75!black!80!yellow}
    }
\newcommand{\wbar}{\overline{\omega}}
\newcommand{\w}{\omega}
\newcommand{\vareps}{\varepsilon}

\newcommand{\F}{\mathcal{F}}

\begin{document}

\title{Preserving fermionic statistics for single-particle approximations in microscopic quantum master equations}

\author{Mikayla Z. Fahrenbruch}
\affiliation{Department of Chemistry, University of Minnesota, Minneapolis, MN 55455 USA}
\author{Anthony W. Schlimgen}
\affiliation{Department of Chemistry, University of Minnesota, Minneapolis, MN 55455 USA}
\author{Kade Head-Marsden}
\email{khm@umn.edu}
\affiliation{Department of Chemistry, University of Minnesota, Minneapolis, MN 55455 USA}
\date{\today}

\begin{abstract}
Microscopic master equations have gained traction for the dissipative treatment of molecular spin and solid-state systems for quantum technologies. Single-particle approximations are often invoked to treat these systems, which can lead to unphysical evolution when combined with master equation approaches. We present a mathematical constraint on the system-environment parameters to ensure that microscopically-derived Markovian master equations preserve fermionic, $N$-representable statistics when applied to reduced systems. We demonstrate these constraints for the recently derived unified master equation and universal Lindblad equation, along with the Redfield master equation for cases when positivity issues are not present. For operators that break the constraint, we explore the addition of Pauli factors to recover $N$-representability. This work promotes feasible applications of novel microscopic master equations for realistic chemical systems.
\end{abstract}

\maketitle

\section{Introduction}
\label{sec:intro}

An open quantum system (OQS) describes a quantum system that interacts with an environment, allowing for exchange of energy or information. OQS approaches can treat a wide range of applications, including photoinduced energy transfer for light-harvesting,~\cite{ishizaki:2009, jeske:2015, storm:2019, cho:2025} quantum optical phenomena of graphene-based nanostructures,~\cite{cox:2014, muller:2020, kosik:2022} and various properties of quantum dots.~\cite{senapati:2025, florian:2016, lettau:2017} There has also been interest in OQS treatments of molecular spin systems for applications in quantum information science, informed by first-principles electronic structure methods or experimental data.~\cite{lunghi:2019,lunghi:2020,lunghi:2022,aruachan:2023,krogmeier:2024, atzori:2016} 

Master equations (MEs) are often used to simulate environmentally-mediated loss over time. Although MEs are equipped for an \textit{N}-electron density matrix, single-particle approximations are often used to limit computational cost when modeling molecules and periodic systems for materials.~\cite{nguyen:2015a, krishtal:2016, topler:2021, lunghi:2020, lunghi:2019, lunghi:2022, cox:2014} One example is the investigation of graphene-based nanomaterials which are relevant for quantum optical technologies.~\cite{kosik:2022, thongrattanasiri:2012, martinez:2022} Even with a tight-binding method to treat just the $p_z$ electrons, a quantum many-body treatment of the dynamics quickly becomes impossible for nanometer-sized flakes.~\cite{pelc:2024} Approximations of the many-body space are thus necessary to capture quantum phenomena of extended solid and condensed phase systems. Such approximations are inherent to quantum-mechanical methods like density functional and Hartree-Fock theories, used to construct the reduced system Hamiltonians.~\cite{kohn:1965, szabo:1996, pelc:2024} However, reduced system dynamics can result in non-physical time evolution for fermions, such as violation of the Pauli-exclusion principle.~\cite{pershin:2008}

Master equations are often categorized as phenomenological, based on macroscopic observations, or microscopic, based on microscopic derivations. The most common phenomenological form is the Gorini-Kossakowski-Sudarshan-Lindblad (GKSL) equation which treats relaxation under the Born-Markov approximation.~\cite{gorini:1976, lindblad:1976,breuer:2010, Manzano:2020} While this ME structure is completely positive and trace preserving (CPTP), violation of the Pauli exclusion principle has been observed using 1-body approximations.~\cite{pershin:2008} Various works have addressed this issue, including the use of \textit{N}-representability conditions and Pauli factors.~\cite{khm:2015, pelc:2024, rosati:2014,rosati:2015, cohen:2009, hod:2016, topler:2021, steinhoff:2012,florian:2016} While the phenomenological approach has its benefits, microscopic MEs are also commonly used, in particular for time-dependent studies of fermionic systems at finite temperatures.~\cite{jeske:2015, lunghi:2019, lunghi:2022} These MEs have the benefit of a more explicit treatment of the environmental degrees of freedom, but come with more challenges for physical representability of the system. Although several studies have explored preserving positivity,~\cite{davies:1974, davies:1976, bacon:1999, lidar:2001, majenz:2013, schaller:2008, trushechkin:2021, farina:2019, cattaneo:2019, cattaneo:2020, mccauley:2020, fernandez:2024, nathan:2020, nathan:2024, litzba:2025, mozgunov:2020, potts:2021} fewer have investigated fermionic, $N$-representable statistics of reduced systems in a microscopic framework.~\cite{zhuang:2020, nguyen:2015, nguyen:2015a} \textcolor{black}{A microscopic non-linear equation has been derived for single-particle fermionic systems, resulting in a single-particle analogue of the Redfield master equation with a secular approximation.~\cite{nguyen:2015} This equation guarantees Fermi-Dirac statistics and has been used to model excited state dynamics of molecular systems.~\cite{nguyen:2015a} }

Here, we present a mathematical constraint on system-environment operators to ensure that the 1-electron reduced density matrix (1-RDM) remains fermionic and $N$-representable after propagation by microscopic MEs. In Sec.~\ref{sec:theory}, we provide the necessary theoretical background, including an overview of three selected microscopic MEs: the Redfield master equation (RME),~\cite{redfield:1957,breuer:2010} unified GKSL master equation (UME),~\cite{trushechkin:2021} and universal Lindblad equation (ULE).~\cite{nathan:2020} We derive \textit{N}-representability constraints for each equation and explore Pauli blocking for operators that break the constraints. In Sec.~\ref{sec:results}, we demonstrate these constraints with population dynamics of a 3-level system and benzene. Although not the primary focus, this work also supports the benchmarking of molecular systems by the ULE and UME, two relatively new MEs. These two master equations present an opportunity to extend the treatment of many-body microscopic dynamics, while preserving physical statistics.

\section{Theory}
\label{sec:theory}

\subsection{Approximate 1-body master equations}
A general master equation takes the form,
\begin{equation}
    \label{eq:me}
    \frac{d\rho(t)}{dt} = -i[H , \rho(t)] + \mathcal{D}(\rho(t)),
\end{equation}
in atomic units, where $\rho(t)$ is the system density matrix, $H $ is the system Hamitonian, $i$ is the imaginary number, $[\cdot, \cdot]$ is the commutator, and $\mathcal{D}(\cdot)$ is the dissipative superoperator in Hilbert space. We can rewrite Eq.~\eqref{eq:me} in superoperator form,
\begin{equation}
    \frac{d\rho(t)}{dt} = \mathcal{L}(\rho(t)),
    \label{eq:superop}
\end{equation}
where $\mathcal{L}(\cdot)$ represents both the Hamiltonian and dissipative components. These types of equations are generally applied to the \textit{N}-body density matrix, which is positive semidefinite, Hermitian, normalized, and antisymmetric with respect to fermionic particle exchange.~\cite{coleman:1963, mazziotti:2012} The size of the Hilbert space grows factorially with the number of electrons, quickly becoming computationally intractable. This exponential scaling can be overcome by considering the 1-RDM, $^1\rho$, \textcolor{black}{which } scales linearly in system size.~\cite{coleman:1963} \textcolor{black}{The 1-RDM is obtained by integrating over all but one electron, which corresponds to tracing out $N-1$ electrons,
\begin{equation}
    {^1\rho} = \Tr_{N-1} |\psi_N\rangle\langle\psi_N|,
\end{equation}
where $\ket{\psi_N}$ is the wavefunction for $N$ fermions. In this work, the 1-electron RDM is trace normalized by the number of electrons, $N$, such that each eigenvalue represents the occupation number of an orbital. We can also express the elements of the 1-RDM as, 
\begin{equation}
    {^1\rho_{ij}} = \langle\psi_N|a_i^\dagger a_j|\psi_N\rangle ,
\end{equation}
where $a_i^\dag$ and $a_j$ are the fermionic creation and annihilation operators for modes $i$ and $j$ respectively. An element of the 1-hole RDM, ${^1q}$, denotes the absence of a particle,
\begin{equation}
    {^1q_{ij}} = \langle\psi_N|a_i a_j^\dagger |\psi_N\rangle.
\end{equation}
The 1-hole RDM is trace normalized by the number of holes, or empty spin orbitals, $r-N$, where $r$ is equal to the total number of spin orbitals.} From the anticommutation relation for fermions, \textcolor{black}{$\{a_j, a_i^\dagger\}=~\delta_{ij}$,~\cite{schilling:2015}} the 1-hole RDM can be expressed as a linear functional of the 1-RDM,
\begin{equation}
    \mathds{1} = {^1\rho} + {^1q},
    \label{eq:pauli}
\end{equation}
where $\mathds{1}$ is the identity. The 1-RDM is \textit{N}-representable if and only if both ${^1\rho}$ and ${^1q}$ are Hermitian, positive semidefinite, and obey Eq.~\eqref{eq:pauli}.~\cite{chakraborty:2014,coleman:1963} For electrons, \textit{N}-representability of the 1-RDM implies the Pauli exclusion principle\textcolor{black}{, meaning that spin-orbital occupations are bound between 0 and 1.} Previous work has considered the implications of these relations in the context of the Lindblad equation, where necessary and sufficient conditions for \textit{N}-representable 1-RDM dynamics were derived.~\cite{khm:2015} Here, we consider the case of a general microscopic master equation that preserves positivity.

\textcolor{black}{To obtain $N$-representability over time, we require that the 1-electron and 1-hole RDMs have complementary dynamics. We investigate superoperators which preserve particle-hole duality by rewriting the 1-electron RDM using Eq.~\eqref{eq:pauli},} 
\begin{equation}
    \label{eq:dqdt}
    \frac{d(\mathds{1}-\leftindex^1q(t))}{dt} = \mathcal{L}(\mathds{1} - \leftindex^{1}q(t)).
\end{equation}
\textcolor{black}{To derive an equation of motion for the 1-hole RDM in the form of Eq.~\eqref{eq:superop},
\begin{equation}
    \label{eq:q-me}
    \frac{d{^1q}(t)}{dt} = \mathcal{L}({^1q}(t)),
\end{equation}
we require,}
\begin{equation}
    \label{eq:constraint}
    \mathcal{L}(\mathds{1}) = 0,
\end{equation}
\textcolor{black}{for a linear map, since the time derivative operates linearly and the identity matrix is independent of time.} \textcolor{black}{To preserve the relationship between $\leftindex^{1}\rho(t)$ and $\leftindex^{1}q(t)$ for all time, $\mathcal{L}(\cdot)$ must correspond to a unital map, meaning that the identity must be conserved as time evolves. Moreover,} if the reduced dynamics are generated by a unital CPTP map, then the particle and hole are guaranteed to each remain positive-semidefinite and have complementary evolutions, \textcolor{black}{therefore ensuring $N$-representability}. \textcolor{black}{The constraint in Eq.~\eqref{eq:constraint} is the key result of this paper. }

\subsection{Microscopic equations and fermionic constraints}

\textcolor{black}{As a concrete demonstration of this constraint, we consider the dynamics governed by three different microscopic MEs. To derive constraints for specific MEs, one can use Eq.~\eqref{eq:pauli} to derive an equation of motion for the 1-electron RDM in terms of the 1-hole RDM to preserve the fermionic anticommutation relation as time evolves. Alternatively, and more simply, we can invoke our general constraint derived in Eq.~\eqref{eq:constraint} by acting the dissipative superoperator on the identity operator and setting the result equal to zero. This allows for constraints on the system-environment coupling parameters to ensure $N$-representability over time. We apply the relation in Eq.~\eqref{eq:constraint} to the RME, UME, and ULE, to analyze the validity of interfacing many-body microscopic master equations with single-particle Hamiltonians.}

\subsubsection{Redfield master equation}

Under the Born-Markov approximation,~\cite{breuer:2010} the Redfield master equation can be written in the Schr\"odinger picture as,
\begin{align}
    \label{eq:rme}
    \dot{\rho}  = -i&[H  + H_\mathrm{LS},\rho ] + \sum_{\w\w\sp{\prime}}\sum_{\alpha\beta} \gamma_{\alpha\beta}(\w,\w\sp{\prime}) \\
    &\times \left(A_{\alpha\w}\rho A^\dag_{\beta\w\sp{\prime}}-\dfrac{1}{2}\{A_{\beta\w\sp{\prime}}^\dag A_{\alpha\w},\rho \}\right) , \notag
\end{align}
where $\gamma_{\alpha\beta}(\w,\w\sp{\prime})$ are decay rates associated with each pair of Bohr frequencies $\w,\w\sp{\prime}$ of the system Hamiltonian, and $\alpha,\beta$ denote different system operators associated with the same Bohr frequency. The Lamb-shift Hamiltonian, $H_\mathrm{LS}$, accounts for a renormalization of the system energy levels due to the environment, and the decay rates and Lamb shifts are derived from a spectral function, $\Gamma_{\alpha\beta}(\w)$. The system-environment coupling operators, $A_{\alpha\w}$, associated with Bohr frequency $\w$, are defined as,
\begin{equation}
    \label{eq:system-operator}
    A_{\alpha\w} = \sum_{jj\sp{\prime}} \Pi_{j}A_\alpha\Pi_{j\sp{\prime}}
\end{equation}
where $\w = \vareps_{j\sp{\prime}}-\vareps_j$ are differences between system eigenvalues, and $\Pi_j = |e_j\rangle\langle e_j |$ are eigenprojectors. Due to the completeness of the system eigenbasis, a summation over all channels returns the original system operator, \textcolor{black}{$A_{\alpha}$.} The system operator is projected into channels that correspond to all possible Bohr frequencies of $H $, and each channel is associated with a microscopically-derived decay rate. This form of the RME requires $A_{\alpha} = A_{\alpha}^\dag$ to enact detailed balance.

\textcolor{black}{Invoking the constraint in Eq.~\eqref{eq:constraint} with the dissipative superoperator defined in the RME yields,} 
\begin{equation}
    \label{eq:general-redfield-constraint}
    0 = \sum_{\w\w\sp{\prime}}\sum_{\alpha\beta} \gamma_{\alpha\beta}(\w,\w\sp{\prime})
    \left[A_{\alpha\w}, A_{\beta\w\sp{\prime}}^\dag\right] ,
\end{equation}
This constraint dictates which system-environment coupling parameters guarantee physical dynamics for multiple electrons in a 1-body framework, provided the populations remain nonnegative. To avoid the risk of negative probabilities, we turn to recently developed microscopic MEs of GKSL form that guarantee positivity. 

\subsubsection{Unified master equation}
\begin{figure}[h]
    \centering
    \includegraphics[width=1\linewidth]{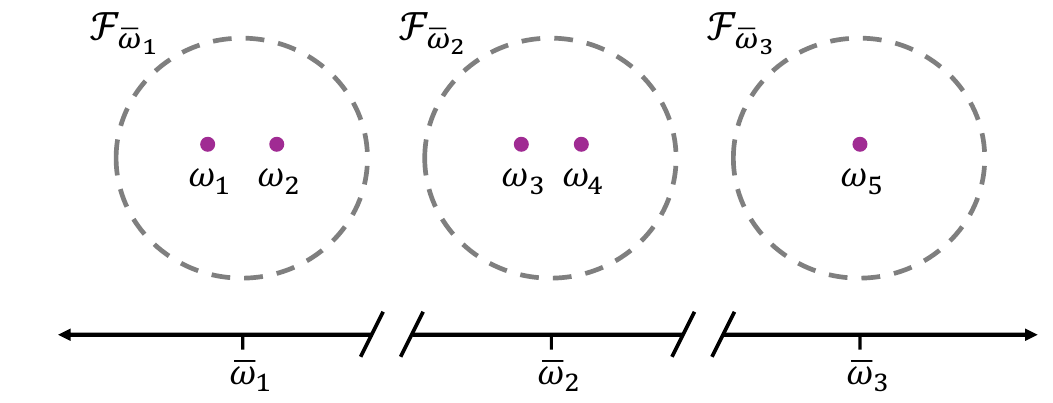}
    \caption{Example of a clustering approach prescribed by the UME with five distinct Bohr frequencies. Note that if all clusters resemble $\F_{\wbar_3}$, the unified master equation reduces to the secular equation.}
    \label{fig:clusters}
\end{figure}
The unified master equation was recently developed to better model near degeneracies while retaining the positivity of the secular, or Davies, equation.~\cite{trushechkin:2021, davies:1974, davies:1976} Its derivation relies on a clustering approach depicted in Figure~\ref{fig:clusters}.~\cite{trushechkin:2021,gerry:2023,vaaranta:2025} The nearby Bohr frequencies are grouped into a cluster $\F_{\wbar}$ such that each cluster is well-separated. Then, each channel associated with $\F_{\wbar}$ is assigned the same decay rate $\gamma_{\alpha\beta}(\wbar)$, where $\wbar$ is the center of the cluster. A secular approximation is applied to the set of averaged frequencies, preserving only the coupling between channels associated with the same cluster; interactions between channels of well-separated frequencies are cleaved. This process results in the UME, 
\begin{align}
    \dot{\rho}  = -i &[H +\Tilde{H}_\mathrm{LS},\rho ] +\sum_{\wbar}\sum_{\w\w\sp{\prime}\in\F_{\wbar}}\sum_{\alpha\beta} \gamma_{\alpha\beta}(\wbar) \\
    &\times \left( A_{\alpha\w}\rho A_{\beta\w\sp{\prime}}^\dag -\frac{1}{2} \{A_{\beta\w\sp{\prime}}^\dag A_{\alpha\w},\rho \} \right) , \notag
\end{align}
presented in the Schr\"odinger picture. The Lamb-shifts and system operators are identical to the RME whereas the decay rates and Lamb-shift Hamiltonian, $\Tilde{H}_\mathrm{LS}$, are now cluster-dependent.~\cite{trushechkin:2021} The UME constraint is,
\begin{equation}
    \label{eq:general-unified-constraint}
     0 = \sum_{\wbar}\sum_{\w,\textcolor{black}{\w\sp{\prime}}\in\F_{\wbar}}\sum_{\alpha\beta} \gamma_{\alpha\beta}(\wbar)
    \left[A_{\alpha\w}, A_{\beta\w\textcolor{black}{\sp{\prime}}}^\dag\right] ,
\end{equation}
and is a condition on channels of the system-environment operator. Since the UME \textcolor{black}{corresponds to a} CPTP \textcolor{black}{map}, this condition guarantees that the dynamics are \textit{N}-representable for the 1-RDM.

\subsubsection{Universal Lindblad equation}
The universal Lindblad equation was introduced around the same time as the UME to achieve positivity without assumptions about spacings between system energy levels.~\cite{nathan:2020} The derivation begins with an integro-differential form of the RME, but a different Markov approximation sends $t_0$ to $-\infty$ rather than 0.~\cite{nathan:2020,nathan:2021,tello:2024} The final result is an equation of GKSL form that induces the same order of magnitude of error as the RME. For a single noise source, the ULE can be written as,
\begin{align}
    \label{eq:ule}
    \dot{\rho}  = -i &[H  + \hat{H}_\mathrm{LS},\rho ] + \sum_{\w\w\sp{\prime}}\sum_{\alpha\beta} \hat{\gamma}_\alpha(\w) \hat{\gamma}^*_\beta(\w\sp{\prime})  \\
    &\times \left(A_{\alpha\w}\rho A^\dag_{\beta\w\sp{\prime}}-\dfrac{1}{2}\{A_{\beta\w\sp{\prime}}^\dag A_{\alpha\w},\rho \}\right), \notag
\end{align}
again in the Schr\"odinger picture. The notation of the original work has been adapted for easier comparison with the UME and RME, and more detail can be found in Appendix~\ref{appendix:ule}. The ULE preserves \textit{N}-representability in the 1-RDM framework when,
\begin{equation}
    \label{eq:general-universal-constraint}
    0 = \sum_{\w,\textcolor{black}{\w\sp{\prime}}}\sum_{\alpha\beta} \hat{\gamma}_{\alpha}(\w) \hat{\gamma}^*_{\beta}(\w\textcolor{black}{\sp{\prime}})
    \left[A_{\alpha\w}, A_{\beta\w\textcolor{black}{\sp{\prime}}}^\dag\right].
\end{equation}
\textcolor{black}{The constraint above is explicitly derived in Appendix~\ref{appendix:ule-constraint-derivation}, and the process is identical for each ME presented in this work.}

\subsection{Pauli blocking \textcolor{black}{as one approach to satisfy $N$-representability}
\label{sec:pauli-blocking}}
There are system operators which violate the \textit{N}-representability constraints but remain necessary to model population transfer; Pauli blocking is one method to address this. A Pauli factor for the $i^\text{th}$ population of the 1-RDM is,
\begin{equation}
    \label{eq:pauli-factor}
    f( {^1\rho}_{ii} ) = 1 -  {^1\rho}_{ii}  ,
\end{equation}
since \textcolor{black}{ $1$ is the maximum occupation of a spin orbital}. When positivity issues are not present, the Pauli factors provide an upper-bound on the eigenvalues of a 1-RDM \textcolor{black}{by closing} population transfer channels once the target states are \textcolor{black}{full}, without compromising the pure dephasing channels.~\cite{pelc:2024, cohen:2009} 

\textcolor{black}{We implement Pauli factors as an ad hoc correction to the three MEs to demonstrate one possible approach for preserving \textit{N}-representability.} \textcolor{black}{Since these Pauli factors are multiplicative and depend on orbital populations, $\leftindex^{1}\rho_{ii}$, they introduce a non-linearity into the MEs.} In spite of the resulting non-GKSL structure, they have previously been shown to preserve positivity.~\cite{rosati:2014} \textcolor{black}{Here, we find that the addition of Pauli factors preserves both positivity and particle-hole duality, and therefore $N$-representability. Similarly, the single-particle analogue of the RME ensures fermionic propagations, and also provides a non-linear mapping.~\cite{nguyen:2015}} \textcolor{black}{More detail on the application of Pauli factors to the RME, UME, and ULE can be found in Appendix~\ref{appendix:pauli-factors}.} 

\section{Results}
\label{sec:results}

\subsection{Benchmark demonstration of constraints}
\label{sec:3lvl-constraint}

To demonstrate the effects of these constraints on the dynamics of a 1-electron RDM, we first consider a three-level ladder system with the following Hamiltonian, 
\begin{align}
    {^1H}  = -0.5|0\rangle\langle 0| + 0|1\rangle\langle 1| + 0.5|2\rangle\langle 2|,
\end{align}
and dissipation governed by the system-environment coupling operator,
\begin{align}
    \label{eq:3lvl-sys-operator}
    A= |0\rangle\langle 1| + |1\rangle\langle 2| + \text{H.c.},
\end{align}
where H.c. denotes the Hermitian conjugate. Since this system lacks near-degeneracies, the model produces identical dynamics for the RME, UME, and ULE. Here, we use the ULE, a CPTP method, to exemplify nuance between dynamics that obey Pauli-exclusion for the 1-particle RDM, and dynamics that are \textit{N}-representable.

At finite temperatures this constraint is not satisfied, \textcolor{black}{shown in Appendix~\ref{appendix:ule-constraint-derivation},} the map is not unital, and therefore the propagated 1-RDM is not \textit{N}-representable. To demonstrate this, we consider dynamics for the initial \textcolor{black}{1-electron RDM},
\begin{align}
    {^1\rho} (0) = |2\rangle\langle 2|,
\end{align}
\textcolor{black}{which corresponds to an initial 1-hole RDM,
\begin{equation}
    {^1q}(0) = |0\rangle\langle0| + |1\rangle\langle1|.
\end{equation}
The dynamics for both ${^1\rho}(t)$ and ${^1q}(t)$} remain positive since the map is CPTP, and the presence of just one electron trivially preserves Pauli-exclusion \textcolor{black}{for the 1-particle RDM.}  \textcolor{black}{For $N$-representability,} the dynamics of the 1-hole RDM must also be considered; the 1-hole RDM and 1-electron RDM must have complementary time evolutions such that Eq.~\eqref{eq:pauli} is satisfied for all time. 

\begin{figure}[h!]
    \centering
    \includegraphics[width=0.95\linewidth]{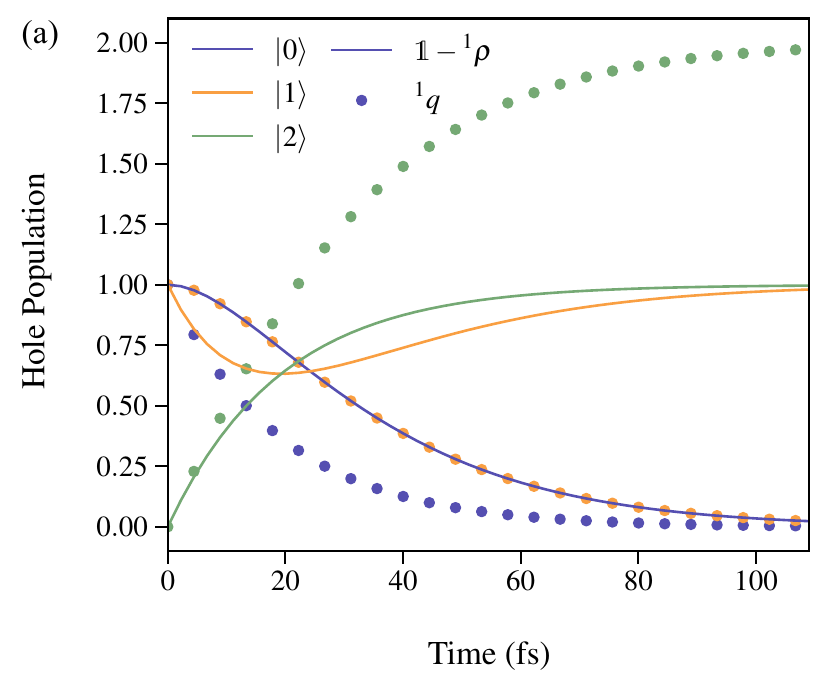}
    \includegraphics[width=0.95\linewidth]{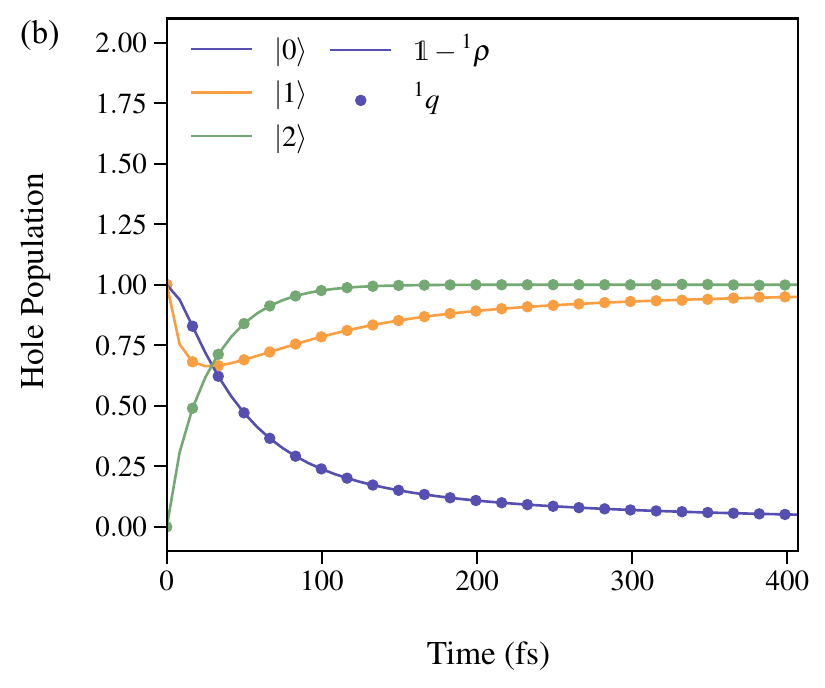}
    \caption{Dynamics of the 1-hole RDM for a 3-level system at 50 K. (a) Application of the ULE, which results in a non-unital map. (b) Application of the Pauli-blocked ULE, which results in a unital map. Both (a) and (b) compare the explicit solution for $\mathds{1} - \leftindex^{1}\rho$, obtained from propagating $\leftindex^{1}\rho$ \textcolor{black}{(lines)}, against the dynamics for $\leftindex^{1}q$ \textcolor{black}{(circles)}.} 
    \label{fig:3lvl-q}
\end{figure}

To investigate the $N$-representability of the resulting \textcolor{black}{state} from the ULE, we can directly evolve $\leftindex^1q(t)$, shown by the filled circles in Figures~\ref{fig:3lvl-q}~(a) and (b). In Figure~\ref{fig:3lvl-q}~(a) we show the non-unital evolution where \textcolor{black}{the constraint} in Eq.~\eqref{eq:constraint}  is not satisfied. \textcolor{black}{The dynamics for $\leftindex^{1}q(t)$ are} positive due to the CP nature of the ULE, \textcolor{black}{but there is} a disagreement \textcolor{black}{between $\leftindex^{1}q(t)$ and} the $N$-representable solution \textcolor{black}{obtained by calculating $\mathds{1} - \leftindex^{1}\rho\textcolor{black}{(t)}$ at each time step, shown by solid lines.}  \textcolor{black}{This emphasizes} that a non-unital CPTP map does not preserve \textit{N}-representability, \textcolor{black}{ since it is unphysical for spin state $|2\rangle$ to be occupied by two holes.}
\textcolor{black}{I}n Figure~\ref{fig:3lvl-q}~(b) we use \textcolor{black}{the Pauli-blocked ULE} for \textcolor{black}{direct propagation} of $\leftindex^{1}q(t)$, \textcolor{black}{shown by circles}. As expected, these results agree with the known $N$-representable dynamics, \textcolor{black}{since the Pauli factors block the ULE from overpopulating state $|2\rangle$}.

\subsection{Benzene}
\label{sec:benzene}
As a molecular example, we consider a benzene system weakly coupled to a bosonic bath with a Drude-Lorentz spectral density. Appendix~\ref{appendix:environment} contains a detailed outline of the spectral functions for each of the MEs, and the same environmental parameters are used throughout Sec.~\ref{sec:results}. In particular, the width of the spectral density is much smaller than the characteristic Bohr frequency to satisfy the Born approximation. 

A geometry optimization and complete active space self-consistent field (CASSCF) calculation with an active space of six electrons and six spatial orbitals were carried out in  ORCA 6.0.1~\cite{roos:1980, roos:2009, siegbahn:1980, siegbahn:1981, neese:2000, neese:2003, neese:2009, neese:2012, neese:2018, neese:2020, neese:2022, neese:2023, kollmar:2019, ugandi:2023, bykov:2015, helmich-paris:2021,izsak:2011, izsak:2012} software for a benzene molecule in the ground state. The B3LYP hybrid functional was used for the geometry optimization, and the def2-TZVP basis set was used for both calculations.~\cite{becke:1993, schafer:1992,schafer:1994,weigend:2005,weigend:2006} 
The system Hamiltonian was constructed from the energies of the CASSCF calculation, 
\begin{align}
    {^1H}  = -0.492|0\rangle\langle0|  -0.323(|1\rangle\langle1| + |2\rangle\langle2|) \\
    + 0.168(|3\rangle\langle3| + |4\rangle\langle4|) + 0.428|5\rangle\langle5| \notag,
\end{align}
in units of Hartrees. We consider an initial reduced system configuration containing two excited electrons,
\begin{equation}
    \label{eq:benzene-initial-state}
    {^1\rho}  = 
    2 |0\rangle\langle 0 | + \sum_{n=1}^4 |n\rangle\langle n |,
\end{equation}
where $|n\rangle$ corresponds to the $n^{\text{th}}$ spatial molecular orbital in the active space. The reduced system operator contains channels for all symmetry-allowed transitions across orbitals,
\begin{align}
    \label{eq:benzene-operator}
    A &= |0\rangle\langle 1| + |0\rangle\langle 2| + |1\rangle\langle 3| + |1\rangle\langle 4| \\ 
    &+ |2\rangle\langle 3| + |2\rangle\langle 4| + |3\rangle\langle 5| + |4\rangle\langle 5| + \text{ H.c.}. \notag
\end{align}

\begin{figure}[h!]
    \centering
    \includegraphics[width=0.95\linewidth]{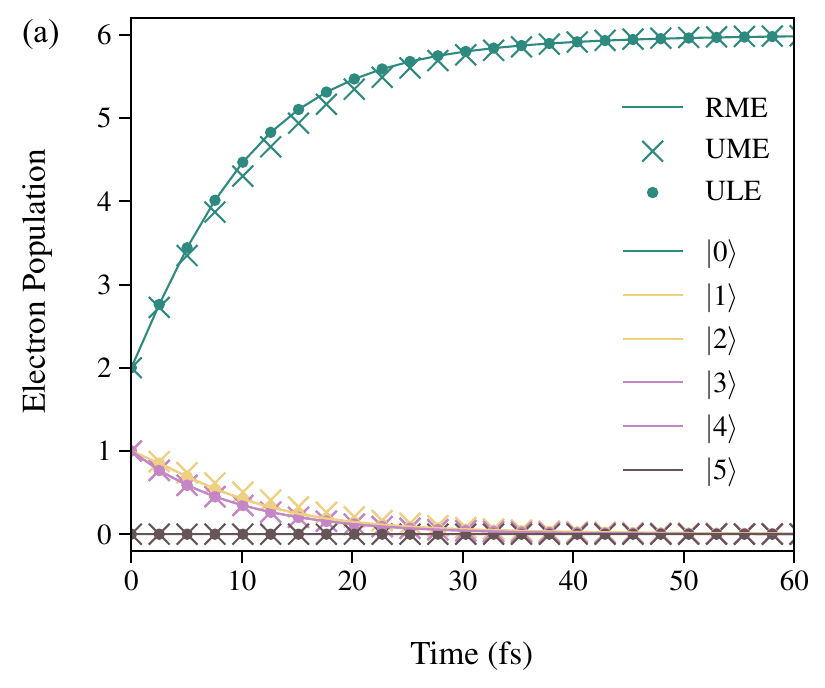}
    \includegraphics[width=0.95\linewidth]{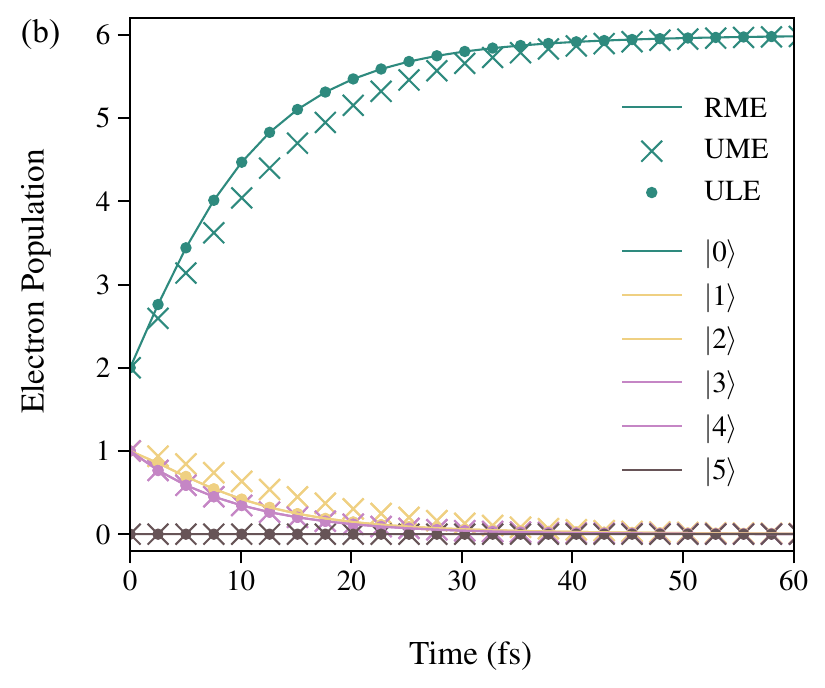}
    \caption{Unconstrained dynamics of benzene at $50$ K by the UME, RME, and ULE that violate the Pauli-exclusion principle. (a) No clustering is invoked in the UME, reducing this equation to the secular or Davies form. (b) A clustering threshold of $0.091$ a.u. is invoked to cluster Bohr frequencies $0.169$ a.u. and $0.260$ a.u.}
    \label{fig:benzene-broken}
\end{figure}

In Figure~\ref{fig:benzene-broken}, we compare  unconstrained dynamics for the RME, UME, and ULE, using different clustering approaches for the UME in (a) and (b), \textcolor{black}{showing the populations of spatial orbitals for simplicity}. In Fig.~\ref{fig:benzene-broken}~(a), no clustering is used such that it reduces to the secular, or Davies, equation, resulting in good agreement with the RME and ULE. In Fig.~\ref{fig:benzene-broken}~(b), we \textcolor{black}{create two clusters: the first containing the transitions $|0\rangle\leftrightarrow|1\rangle$ , $|0\rangle\leftrightarrow|2\rangle$, $|3\rangle\leftrightarrow|5\rangle$, and $|4\rangle\leftrightarrow|5\rangle$, and the second containing the transitions $|1\rangle\leftrightarrow|3\rangle$ , $|1\rangle\leftrightarrow|4\rangle$, $|2\rangle\leftrightarrow|3\rangle$, and $|2\rangle\leftrightarrow|4\rangle$. Noting that the transitions in the second cluster are degenerate, while the transitions in the first cluster were grouped using a  threshold of 0.091 a.u.} In this case, the dynamics governed by the UME are slightly slower than the RME and ULE, suggesting that this clustering is not ideal for a spectral density width of $0.01$ a.u.. This highlights that the determination of clustering thresholds can be nontrivial for practical implementation of the UME. 

As shown in both Fig.~\ref{fig:benzene-broken}~(a) and (b), naive application of the RME, UME, and ULE drives six electrons into the same energy level at 50 K. This is unphysical for fermions and is consistent with violation of the constraints \textcolor{black}{described explicitly in Appendix~\ref{appendix:benzene-constraints}}. Since detailed balance exponentially damps spectral functions for $\w<0$, the chosen system operator does not ensure a fermionic evolution for any of the three MEs at finite temperatures\textcolor{black}{, as discussed in Appendix~\ref{appendix:benzene-constraints}}. In general, there are no population transfer operators for this system that guarantee an \textit{N}-representable state and simultaneously obey detailed balance. This is a key difference from previous work done in the context of Lindblad dynamics.~\cite{khm:2015}

\begin{figure}[h!]
    \centering
    \includegraphics[width=0.95\linewidth]{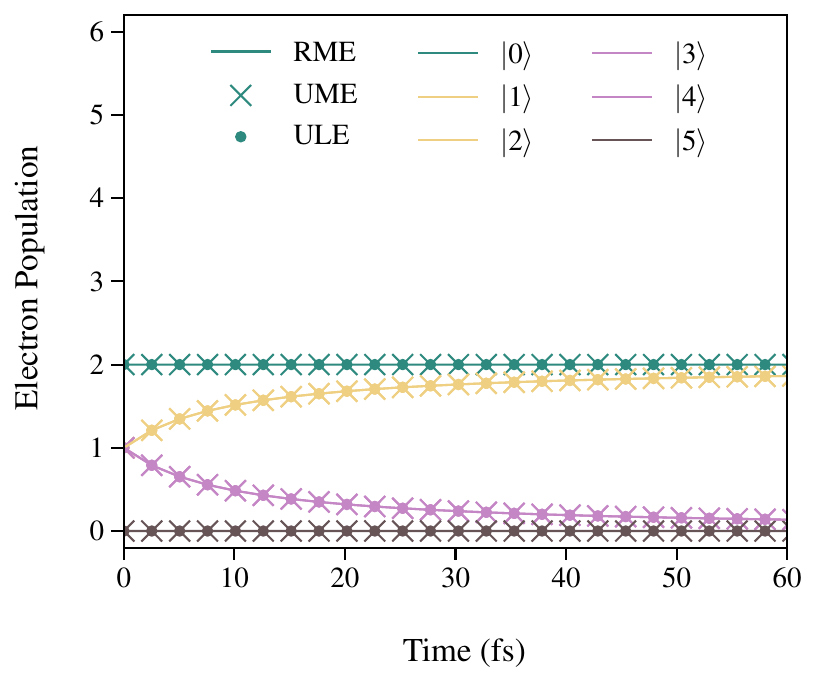} \caption{Use of Pauli blocking to recover fermionic statistics for benzene with the RME, UME, and ULE at 50 K.}
    \label{fig:benzene-pauli-blocked}
\end{figure}

To recover \textit{N}-representable dynamics, one solution is the application of Pauli blocking, shown in Figure~\ref{fig:benzene-pauli-blocked}. \textcolor{black}{The Pauli-blocked RME, UME, and ULE all preserve positivity and particle-hole duality, resulting in $N$-representable dynamics}. The Pauli-blocked dynamics exhibit asymptotic behavior due to the Pauli factors, which slow relaxation as the target state becomes fully occupied.~\cite{nguyen:2015}

\section{Discussion and conclusions}
\label{sec:disc}
Along with the widely used RME, this work explores the UME and ULE to treat fermionic dynamics in a 1-body framework common for molecular and material applications. As demonstrated for the selected equations, direct implementation of microscopic MEs under this approximation can be problematic for modeling population transfer. We present a constraint on linear reduced system Liouvillians that is necessary and sufficient to guarantee \textit{N}-representability of the propagated 1-RDM, assuming that positivity is preserved. This constraint is \textcolor{black}{equivalent to the resulting map being unital}. Although not explicitly discussed, pure dephasing operators \textcolor{black}{generally} preserve \textit{N}-representability due to their diagonal forms. 

\textcolor{black}{Due to detailed balance, CPTP microscopic MEs that incorporate population transfer operators generally do not satisfy the unital mapping constraint at finite temperatures. While a general fix is beyond the scope of the present work, one possible approach for recovering $N$-representable dynamics is the addition of Pauli-blocking factors to the MEs. We note that the self-consistent, single-particle analogue of the RME also ensures fermionic statistics.~\cite{nguyen:2015} Although our constraint was derived for linear MEs, we find that the Pauli-blocked MEs preserve both positivity and particle-hole duality, thus preserving the $N$-representability of the system.} 

The constraints presented here highlight that the traditional many-body forms of microscopic master equations are not inherently designed for reduced fermionic systems. Even when the Pauli-exclusion principle is not violated for the 1-\textcolor{black}{electron} RDM, a non-unital CPTP map will not provide a \textcolor{black}{complementary} evolution for the 1-hole RDM. Care needs to be taken when interfacing microscopic equations with single-particle approximations, such as DFT and Hartree-Fock, for many-electron systems. In general, this work formally establishes a foundation for feasible and physical modeling of reduced fermionic systems by extending the many-body microscopic framework for single-particle approximations.

\section{Acknowledgements}
KHM acknowledges the start-up funds from the University of Minnesota and the Minnesota Supercomputing Institute.

\section{Data Availability Statement}
The data that support the findings of this study are available from the corresponding author upon reasonable request.

\renewcommand*{\bibfont}{\normalsize}
\section{References}
\bibliography{main}

\begin{appendix}
\section{More details on the universal Lindblad equation}
\label{appendix:ule}

We have adapted the form of the ULE presented in Eq.~\eqref{eq:ule} from the original work in Ref.~[\!\citenum{nathan:2020}] for consistency across the three explored microscopic MEs. The coupling parameter, defined as $\gamma$ in the original derivation, is reabsorbed into the bath operator to be addressed with a spectral density.~\cite{nathan:2020} The Lamb-shift Hamiltonian and decay rates in Eq.~\eqref{eq:ule} are denoted with a hat. The decay rates are defined as,
\begin{equation}
    \hat{\gamma}_\alpha(\w) = \sqrt{2\pi\hat{\Gamma}_{\alpha\alpha}(\w)} ,
\end{equation}
where $\hat{\Gamma}_{\alpha\beta}(\w)$ corresponds to the bath spectral function. Instead of the one-sided Fourier transform (FT) of the bath correlation function, $C_{\alpha\beta}(t)$, used by the RME and UME, a traditional FT is used,
\begin{equation}
    \label{eq:ule-spectral-function}
    \hat{\Gamma}_{\alpha\beta}(\w)= \dfrac{1}{2\pi}\int_{-\infty}^\infty dt\, e^{i\w t}C_{\alpha\beta}(t) .
\end{equation}
The Lamb-shift Hamiltonian now involves a triple summation over the eigenvalues of $H $,
\begin{equation}
    \hat{H}_\mathrm{LS} = \sum_{lmn}\sum_{\alpha\beta}  \hat{S}_{\alpha\beta}(\w_{ml},\w_{ln})A_{\alpha\w_{ml}} A_{\beta\w_{ln}} ,
\end{equation}
such that $\w_{ml} = \vareps_l - \vareps_m$, along with a different method to calculate the Lamb shifts,
\begin{align}
    \hat{S}_{\alpha\beta}(\w_{ml},\w_{ln}) &= -2\pi\,\mathcal{P}\int_{-\infty}^{\infty} d\w \,\w^{-1} 
    \mathcal{Q}(\w_{ml},\w_{ln}) ,\\
    \mathcal{Q}(\w_{ml},\w_{ln}) &= \sqrt{\hat{\Gamma}_{\alpha\beta}(\w-\w_{ml})\hat{\Gamma}_{\alpha\beta}(\w+\w_{ln})} ,
\end{align}
where $\mathcal{P}$ indicates the Cauchy principal value integral.
\textcolor{black}{\section{Constraint for universal Lindblad equation}
\label{appendix:ule-constraint-derivation}}
\textcolor{black}{As a concrete example of applying the constraint, we provide an explicit constraint derivation for the ULE. The ULE is a CP map, meaning that both the 1-electron RDM and 1-hole RDM remain positive semidefinite for all time,
\begin{align}
    {^1\rho}(t) &\succeq 0 ,\\
    {^1q}(t) &\succeq 0 .
\end{align}
Therefore satisfying Eq.~\eqref{eq:constraint} guarantees that ${^1\rho}(t)$ will remain \textit{N}-representable for all time. We apply the constraint to the ULE by examining the propagation of the identity matrix,
\begin{align}
    \label{eq:ule-for-identity}
    \mathcal{L}(\mathds{1})  = -i &[H  + \hat{H}_\mathrm{LS},\mathds{1} ] + \sum_{\w\w\sp{\prime}}\sum_{\alpha\beta} \hat{\gamma}_\alpha(\w) \hat{\gamma}^*_\beta(\w\sp{\prime})  \\
    &\times \left(A_{\alpha\w}\mathds{1} A^\dag_{\beta\w\sp{\prime}}-\dfrac{1}{2}\{A_{\beta\w\sp{\prime}}^\dag A_{\alpha\w},\mathds{1} \}\right) \notag
\end{align}
which simplifies to,
\begin{equation}
    \label{eq:appendix-ule-constraint}
    0 = \sum_{\w\w\sp{\prime}}\sum_{\alpha\beta} \hat{\gamma}_\alpha(\w) \hat{\gamma}^*_\beta(\w\sp{\prime})  
    \left[A_{\alpha\w},A^\dag_{\beta\w\sp{\prime}}\right].
\end{equation}
To demonstrate that the constraint is not satisfied in Sec.~\ref{sec:3lvl-constraint}, we apply the constraint to the 3-level system and find that it is not satisfied at 50 K, 
\begin{align}
    0\not=3(|\hat{\gamma}(0.5)|^2-|\hat{\gamma}(-0.5)|^2)(|0\rangle\langle0|-|2\rangle\langle2|) ,
\end{align}
since the spectral function is not symmetric at finite temperature, as required by detailed balance.} 

\textcolor{black}{\section{Pauli factors in microscopic master equations}
\label{appendix:pauli-factors}}

\textcolor{black}{We use Pauli factors to provide a concrete example of population dynamics which guarantee \textit{N}-representability. Since Pauli blocking preserves Hamiltonian evolution, only the Hilbert space dissipator, $\mathcal{D}({^1\rho} )$, is considered for each equation, and we use the following notation for the original dissipator in each equation,
\begin{equation}
    {T}_{ijlk}^{\alpha\beta} = A_{\alpha\w_{ij}}{^1\rho}  A^\dag_{\beta\w_{lk}}-\dfrac{1} 
    {2}\{A_{\beta\w_{lk}}^\dag A_{\alpha\w_{ij}}, {^1\rho} \}.
\end{equation}
In order to prevent overpopulation of a state $|i\rangle$, we multiply ${T}_{ijlk}^{\alpha\beta}$ by the Pauli factor $f({^1\rho}_{ii})$ for $i\not=j$. To preserve pure dephasing channels, dissipator terms with $i=j$ are not multiplied by Pauli factors, similar to previous work.~\cite{pelc:2024}} \textcolor{black}{The Pauli-blocked dissipators for the RME, UME, and ULE are,
\begin{align}
    \label{eq:rme-adaptation}
    \mathcal{D}({^1\rho} ) = \sum_{i\not=j,l\not=k}\sum_{\alpha\beta} &f( {^1\rho}_{ii} ) \gamma_{\alpha\beta}(\w_{ij},\w_{lk}){T}_{ijlk}^{\alpha\beta}\\
    &\quad+\gamma_{\alpha\beta}(0,0){T}_{iill}^{\alpha\beta},\notag
\end{align}
\begin{align}
    \label{eq:ume-adaptation}
    \mathcal{D}({^1\rho} ) &= \sum_{\wbar\not=0} \sum_{ijlk}\sum_{\alpha\beta} f( {^1\rho}_{ii} ) \gamma_{\alpha\beta}(\wbar){T}_{ijlk}^{\alpha\beta}\\ 
    &\quad+\sum_{\wbar=0} \sum_{ijlk}\sum_{\alpha\beta} \gamma_{\alpha\beta}(0){T}_{ijlk}^{\alpha\beta} ,\notag 
\end{align}
such that $\w_{ij},\w_{lk}\in\F_{\wbar}$ and,
\begin{align}
    \label{eq:ule-adaptation}
    \mathcal{D}({^1\rho} ) =  \sum_{i\not=j,l\not=k}\sum_{\alpha\beta} &f( {^1\rho}_{ii} ) {\hat{\gamma}}_{\alpha}(\w_{ij}) {\hat{\gamma}}^*_{\beta}(\w_{lk}) {T}_{ijlk}^{\alpha\beta} \\
    &\quad+\hat{\gamma}_{\alpha}(0){\hat{\gamma}}^*_{\beta}(0){T}_{iill}^{\alpha\beta},\notag
\end{align}
respectively.}
\textcolor{black}{The Pauli factors used in this paper are effectively a phenomenological correction which preserves the relationship between the 1-particle and 1-hole RDMs over time and thus results in $N$-representable dynamics.}

\section{Treatment of a bosonic bath}
\label{appendix:environment}
This work uses a bosonic bath with infinitely many states available and a  Hamiltonian of the following form,
\begin{equation}
    H_\mathrm{B} = \sum_{k} \w_k b_k^\dag b_k,
\end{equation}
where $k$ is a bath mode and $b_k$ ($b_k^\dag$) is a bosonic annihilation (creation) operator. The bath is assigned an operator,
\begin{equation}
    B = \sum_k g_k (b_k^\dag + b_k),
\end{equation}
where $g_k$ is the effective coupling constant between a system and bath mode in the interaction Hamiltonian $H_\mathrm{I}$. Since there is only one bath in this work, the correlation function becomes an autocorrelation function,
\begin{align}
    \label{eq:expanded-bcf}
    C(t) = \sum_{jk} g_k^*g_j
    \left[\langle b_k b_j^\dag \rangle e^{-i\w_k t} + \langle b_k b_j \rangle e^{-i\w_k t} \right. \\
    \left. +\langle b_k^\dag b_j^\dag \rangle e^{i\w_k t} + \langle b_k^\dag b_j \rangle e^{i\w_k t}  \right]  \notag ,
\end{align}
where $\langle b_k b_j \rangle$ is a thermal average between modes $k$ and $j$ in the bath. Note that two of the thermal averages can be rewritten via the definition of a number operator $N_k$,
 \begin{align}
     \langle b_k^\dag b_j \rangle &= \delta_{jk} N_k, \\
     \langle b_k b_j^\dag \rangle &= \delta_{jk} (N_k +1), 
 \end{align}
 where $\delta_{jk}$ is the Kronecker delta. The number operator counts how many bosons occupy a given state, and its average value can be calculated with what has been referred to as either the Planck distribution~\cite{breuer:2010} or the Bose-Einstein distribution,~\cite{gerry:2023}
\begin{equation}
    N(\w_k) = \dfrac{1}{\exp(\hbar\w_k/k_\mathrm{B}T)-1} ,
\end{equation}
where $k_\mathrm{B}$ is the Boltzmann constant and $\hbar$ is the reduced Planck constant, in atomic units here. After evaluation of the thermal averages in Eq.~\eqref{eq:expanded-bcf} and application of a continuum limit, the bath correlation function simplifies to,
 \begin{equation}
     C(t) \!=\! \int_{0}^{\infty} \!\!\! d\w J(\w) \! \left[(N(\w)+1)e^{-i\w t} \!+\! N(\w)e^{i\w t}\right] \!,
 \end{equation}
where $J(\w)$ is called a spectral density and approximates $|g_k|^2$ to quantify system-environment coupling.
It is now possible to determine the spectral function for each of the master equations. Starting with the RME and UME, $\Gamma(\w_0)$ is the one-sided FT of the bath correlation function, which can be treated with the following relation, 
\begin{equation}
    \label{eq:cauchy}
    \int_0^\infty ds e^{-i x s} = \pi \delta(x) - i \mathcal{P} \frac{1}{x}.
\end{equation}
The Cauchy principal value is again denoted as $\mathcal{P}$, and $\delta (x)$ is the Dirac delta function. Application of Eq.~\eqref{eq:cauchy} results in the spectral function for the RME and UME 
below,
\begin{align} 
    \label{eq:Gamma-rme/ume}
    \Gamma(\w_0)=
    \begin{cases}
    \pi J(\w_0)(N(\w_0)+1) + \xi &  \w_0 > 0 \\
    \pi J(-\w_0)N(-\w_0) +  \xi &  \w_0 < 0 \\
    \lim_{\w_0\to 0} \pi J(\w_0)(2N(\w_0)+1) + \xi & \w_0 = 0
    \end{cases}.
\end{align}
The principal value integral $\xi$ has the following form,
\begin{align}
    \xi = \mathcal{P} \int_0 ^\infty d\w J(\w) &\left[N(\w) \left(\dfrac{i}{\w_0 +\w}\right) \right. \\
    &\left. +(N(\w) +1) \left(\dfrac{i}{\w_0 - \w}\right)\right] \notag,
\end{align}
and is evaluated numerically for each case in Eq~\eqref{eq:Gamma-rme/ume}. Deriving the spectral function for the ULE mirrors the same process, except it is now defined as the traditional FT of the bath correlation function. The integration bounds of $(-\infty,\infty)$ in Eq.~\eqref{eq:ule-spectral-function} call for the use of the following relation,
\begin{equation}
    \int_{-\infty}^\infty ds e^{-i x s} = 2\pi \delta(x), 
\end{equation}
where the Cauchy principal values are no longer involved. The three possible cases for the ULE spectral function are,
\begin{align} 
    \label{eq:Gamma-ule}
    \hat{\Gamma}(\w_0)=
    \begin{cases}
    J(\w_0)(N(\w_0)+1) &  \w_0 > 0 \\
    J(-\w_0)N(-\w_0) &  \w_0 < 0 \\
    \lim_{\w_0\to 0}J(\w_0)(2N(\w_0)+1) & \w_0 = 0
    \end{cases}.
\end{align}
All that is left is the substitution of an appropriate spectral density for a working form of the spectral functions in Eq.~\eqref{eq:Gamma-rme/ume} and Eq.~\eqref{eq:Gamma-ule}.

\begin{figure}[H]
    \centering
    \includegraphics[width=1\linewidth]{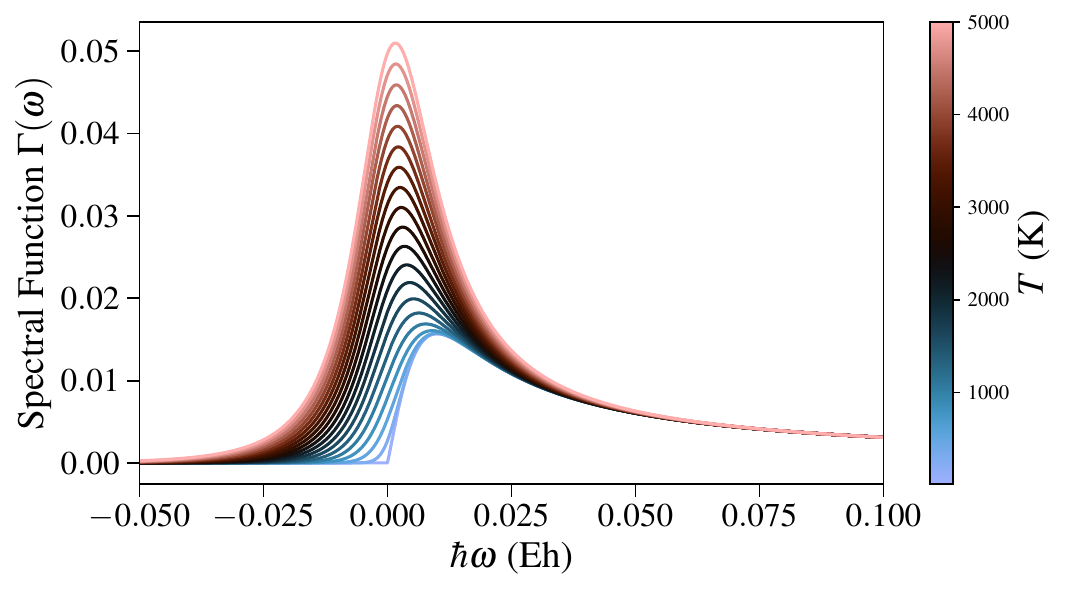}
    \caption{Spectral function $\Gamma(\w)$ for a bosonic bath with a Drude-Lorentz spectral density with $\lambda = 0.01$ a.u. and $\hbar$ in atomic units. The color map denotes the temperature range.}
    \label{fig:spectral-function}
\end{figure}

This work utilizes a Drude-Lorentz spectral density,
\begin{equation}
    J(\w_0)=\frac{\w_0\lambda^2}{\w_0^2 +\lambda^2},
\end{equation}
where $\lambda$ is a parameter chosen to dictate the width of the spectral density.  The spectral function used for the UME and RME is depicted in Figure~\ref{fig:spectral-function}, which is identical to that of the ULE beyond the scaling of the $y$-axis. 

\textcolor{black}{\section{Constraints for benzene}
\label{appendix:benzene-constraints}}
\textcolor{black}{As a molecular example, we apply the \textit{N}-representability constraint to assess the reduced dynamics of benzene.
We evaluate Eq.~\eqref{eq:general-redfield-constraint} with the system operator from Eq.~\eqref{eq:benzene-operator} to solve for the evolution of the identity with the RME,
\begin{align}
    \label{eq:benzene-constraint-rme}
    \mathcal{L}(\mathds{1}) &= 6(2(\gamma(\w_{01})-\gamma(\w_{10}))|0\rangle\langle0| \\
    &+ 2(\gamma(\w_{13},\w_{10})-\gamma(\w_{01},\w_{31}))(|0\rangle\langle3|+|0\rangle\langle4|) \notag \\
    &+ 2(\gamma(\w_{31},\w_{01})-\gamma(\w_{10},\w_{13}))(|3\rangle\langle0|+|4\rangle\langle0|) \notag \\
    &+ 2(\gamma(\w_{13},\w_{53})-\gamma(\w_{35},\w_{31}))(|1\rangle\langle5|+|2\rangle\langle5|) \notag \\
    &+ 2(\gamma(\w_{53},\w_{13})-\gamma(\w_{31},\w_{35}))(|5\rangle\langle1|+|5\rangle\langle2|) \notag \\
    &+ (\gamma(\w_{10})-\gamma(\w_{01}) + 2\gamma(\w_{13})-2\gamma(\w_{31})) \notag \\  &\quad\times(|1\rangle\langle1|+|1\rangle\langle2|+|2\rangle\langle1|+|2\rangle\langle2|) \notag \\
    &+ (\gamma(\w_{35})-\gamma(\w_{53}) + 2\gamma(\w_{31})-2\gamma(\w_{13})) \notag \\  &\quad\times(|3\rangle\langle3|+|3\rangle\langle4|+|4\rangle\langle3|+|4\rangle\langle4|) \notag \\
    &+ 2(\gamma(\w_{53})-\gamma(\w_{35}))|5\rangle\langle5|), \notag
\end{align}
where $\gamma(\w)$ denotes $\gamma(\w,\w)$ for simplicity. Similarly, we expand the constraint in Eq.~\eqref{eq:general-unified-constraint} for the UME,
\begin{align}
    \label{eq:benzene-constraint-ume}
    \mathcal{L}(\mathds{1}) &=2 \left(\gamma(\w_{01}) - \gamma(\w_{10})\right) |0\rangle\langle0| \\
    &+\left(- \gamma(\w_{01}) + \gamma(\w_{10}) + 2 \gamma(\w_{13}) - 2 \gamma(\w_{31})\right) |1\rangle\langle1| \notag\\
    &+\left(- \gamma(\w_{01}) + \gamma(\w_{10}) + 2 \gamma(\w_{13}) - 2 \gamma(\w_{31})\right)  |2\rangle\langle2|\notag\\
    &+ \left(- 2 \gamma(\w_{13}) + 2 \gamma(\w_{31}) + \gamma(\w_{35}) - \gamma(\w_{53})\right) |3\rangle\langle3|  \notag\\
     &+ \left(- 2 \gamma(\w_{13}) + 2 \gamma(\w_{31}) + \gamma(\w_{35}) - \gamma(\w_{53})\right) |4\rangle\langle4|  \notag\\
    &+ 2 \left(- \gamma(\w_{35}) + \gamma(\w_{53})\right) |5\rangle\langle5|. \notag
\end{align}
Lastly, we apply Eq.~\eqref{eq:general-universal-constraint} to examine the ULE,
\begin{align}
    \label{eq:benzene-constraint-ule}
    \mathcal{L}(\mathds{1}) &= 6(2(|\hat{\gamma}(\w_{01})|^2-|\hat{\gamma}(\w_{10})|^2)|0\rangle\langle0| \\
    &+ 2(\hat{\gamma}(\w_{13})\hat{\gamma}^*(\w_{10})-\hat{\gamma}(\w_{01})\hat{\gamma}^*(\w_{31}))(|0\rangle\langle3|+|0\rangle\langle4|) \notag \\
    &+ 2(\hat{\gamma}(\w_{31})\hat{\gamma}^*(\w_{01})-\hat{\gamma}(\w_{10})\hat{\gamma}^*(\w_{13}))(|3\rangle\langle0|+|4\rangle\langle0|) \notag \\
    &+ 2(\hat{\gamma}(\w_{13})\hat{\gamma}^*(\w_{53})-\hat{\gamma}(\w_{35})\hat{\gamma}^*(\w_{31}))(|1\rangle\langle5|+|2\rangle\langle5|) \notag \\
    &+ 2(\hat{\gamma}(\w_{53})\hat{\gamma}^*(\w_{13})-\hat{\gamma}(\w_{31})\hat{\gamma}^*(\w_{35}))(|5\rangle\langle1|+|5\rangle\langle2|) \notag \\
    &+ (|\hat{\gamma}(\w_{10})|^2-|\hat{\gamma}(\w_{01})|^2 + 2|\hat{\gamma}(\w_{13})|^2-2|\hat{\gamma}(\w_{31})|^2) \notag \\  &\quad\times(|1\rangle\langle1|+|1\rangle\langle2|+|2\rangle\langle1|+|2\rangle\langle2|) \notag \\
    &+ (|\hat{\gamma}(\w_{35})|^2-|\hat{\gamma}(\w_{53})|^2 + 2|\hat{\gamma}(\w_{31})|^2-2|\hat{\gamma}(\w_{13})|^2) \notag \\  &\quad\times(|3\rangle\langle3|+|3\rangle\langle4|+|4\rangle\langle3|+|4\rangle\langle4|) \notag \\
    &+ 2(|\hat{\gamma}(\w_{53})|^2-|\hat{\gamma}(\w_{35})|^2)|5\rangle\langle5|). \notag
\end{align}
These three equations are, in general, not equal to zero at 50~K, and therefore do not satisfy the necessary constraint for $N$-representability.}
\clearpage
\end{appendix}

\end{document}